%% file: main.tex
\documentclass[a4,journal,final]{IEEEtran}
\usepackage{algorithmic}
\usepackage[utf8]{inputenc} 
\usepackage[T1]{fontenc}    
\usepackage[pagebackref=true,breaklinks=true,bookmarks=false]{hyperref}      
\usepackage{booktabs} 
\usepackage{amsfonts}       
\usepackage{nicefrac}       
\usepackage{microtype}      
\usepackage[accsupp]{axessibility}
\usepackage{float}
\usepackage{amssymb}
\usepackage{enumitem}
\usepackage{algorithm}
\usepackage[algo2e]{algorithm2e}
\usepackage{amsmath,amsthm,mathtools}
\usepackage{url}
\usepackage{stmaryrd}
\usepackage{cite}
\usepackage{multirow}
\usepackage[table,xcdraw]{xcolor}
\usepackage{threeparttable}
\usepackage{graphicx}
\usepackage{doi}
\usepackage{mathptmx}
\usepackage{amssymb}
\usepackage{subfig}
\usepackage{caption}
\usepackage{array}
\usepackage{tabularx}

\setlist{nolistsep}

\def\BibTeX{{\rm B\kern-.05em{\sc i\kern-.025em b}\kern-.08em
    T\kern-.1667em\lower.7ex\hbox{E}\kern-.125emX}}

\DeclareMathOperator{\diag}{\operatorname{diag}}

\begin{document}

\title{Time-varying Signals Recovery via Graph Neural Networks}

\author{\textit{John A. Castro-Correa$^{1*}$\thanks{*\textbf{Authors have equal contributions}} \thanks{Corresponding author: jcastro@udel.edu}, Jhony H. Giraldo$^{2*}$, Anindya Mondal$^{3*}$}, \\ \textit{Mohsen Badiey$^1$, Thierry Bouwmans$^4$, Fragkiskos D. Malliaros$^5$} \\
$^1$ University of Delaware, USA\\
$^2$ LTCI, Télécom Paris, Institut Polytechnique de Paris, France \\
$^3$ Jadavpur University, India \\
$^4$ Laboratoire Mathématiques, Image et Applications (MIA), La Rochelle Université, France \\
$^5$ Université Paris-Saclay, CentraleSupélec, Inria, Centre for Visual Computing (CVN), France \\
}

\markboth{Accepted in IEEE ICASSP 2023}{}

\maketitle

\begin{abstract}
The recovery of time-varying graph signals is a fundamental problem with numerous applications in sensor networks and forecasting in time series. Effectively capturing the spatio-temporal information in these signals is essential for the downstream tasks. Previous studies have used the smoothness of the temporal differences of such graph signals as an initial assumption. Nevertheless, this smoothness assumption could result in a degradation of performance in the corresponding application when the prior does not hold. In this work, we relax the requirement of this hypothesis by including a learning module. We propose a Time Graph Neural Network (TimeGNN) for the recovery of time-varying graph signals. Our algorithm uses an encoder-decoder architecture with a specialized loss composed of a mean squared error function and a Sobolev smoothness operator.TimeGNN shows competitive performance against previous methods in real datasets. 
\end{abstract}


\section{Introduction}

\input{tex/intro}

\section{Time Graph Neural Network}\label{TimeGNN}

\input{tex/timegnn}

\section{Experiments and Results}\label{experiments_results}

\input{tex/experiements}

\section{Conclusions}

\input{tex/conclusions}\label{conclusions}

\end{document}

%% file: tex/intro.tex
Recent advances in information technology have led to an accumulation of large amounts of unstructured data.
The representation and analysis of such irregular and complex data is a daunting task.
Graph Signal Processing (GSP) and Graph Neural Networks (GNNs) are emerging research fields that have proved to be helpful for such tasks in recent years \cite{1,2,3,4,5,6}.
In GSP and GNNs, the data is modeled as signals or vectors on a set of nodes of a graph, incorporating both the feature information and the underlying structure of the data.
GSP and GNNs thus provide new perspectives on data handling, connecting machine learning and signal processing \cite{7}, with profound impact in various fields like semi-supervised learning \cite{3}, computer vision \cite{8,9}, and social media \cite{10}.

The sampling and reconstruction of graph signals are fundamental tasks that have recently attracted considerable attention from the signal processing and machine learning communities \cite{11,12,13,1,14,15,16,17,18}.
Nevertheless, the problem of time-varying graph signal reconstruction\footnote{One can think of the recovery of time-varying graph signals as a matrix completion task where each column (or row) is associated with time, and each row (or column) is associated with a vertex of a graph.} has not been widely explored \cite{18}.
The reconstruction of time-varying graph signals has significant applications in data recovery in sensor networks, forecasting of time-series, and infectious disease prediction \cite{16,20,21,22,18}.
Previous studies have extended the definition of smooth signals from static to time-varying graph signals \cite{23}.
Similarly, other works have focused on the rate of convergence of the optimization methods used to solve the reconstruction problem \cite{18,20}.
However, the success of these optimization-based methods requires appropriate prior assumptions about the underlying time-varying graph signals, which could be inflexible for real-world applications.

In this work, we propose the Time Graph Neural Network (TimeGNN) model to recover time-varying graph signals.
TimeGNN encodes the time series of each node in latent vectors.
Therefore, these embedded representations are decoded to recover the original time-varying graph signal.
Our architecture comprises: 1) a cascade of Chebyshev graph convolutions \cite{2} with increasing order and 2) linear combination layers.
Our algorithm considers spatio-temporal information using: 1) graph convolutions \cite{2} and 2) a specialized loss function composed of a Mean Squared Error (MSE) term and a Sobolev smoothness operator \cite{18}.
TimeGNN shows competitive performance against previous methods in real-world datasets of time-varying graph signals.

The main contributions of our work are summarized as follows: 1) we exploit GNNs to recover time-varying graph signals from their samples, 2) we relax the strict prior assumption of previous methods by including some learnable modules in TimeGNN, and 3) we perform experimental evaluations on natural and artificial data, and compare TimeGNN to four methods of the literature.
The rest of the paper is organized as follows.
Section \ref{TimeGNN} introduces the proposed TimeGNN model.
Section \ref{experiments_results} presents the experimental framework and results.
Finally, Section \ref{conclusions} shows the conclusions.

%% file: tex/timegnn.tex
\subsection{Preliminaries}
\label{sec:math}

We represent a graph with $G = (\mathcal{V}, \mathcal{E}, \mathbf{A})$, where $\mathcal{V}$ is the set of nodes with $\vert \mathcal{V} \vert=N$, ${\mathcal{E}\subseteq \{(i,j)\mid i,j\in \mathcal{V}\;{\textrm {and}}\;i\neq j\}}$ is the set of edges, and $\mathbf{A} \in \mathbb{R}^{N \times N}$ is the weighted adjacency matrix with $\mathbf{A}(i,j)=a_{i,j} \in \mathbb{R}_+$ if $(i,j) \in \mathcal{E}$ and $0$ otherwise.
In this work, we consider connected, undirected, and weighted graphs.
We also define the symmetrized Laplacian as 
$\mathbf{L} = \mathbf{I}- \mathbf{D}^{-\frac{1}{2}}\mathbf{A}\mathbf{D}^{-\frac{1}{2}}$
, where 

$\mathbf{D}=diag(\mathbf{A1})$ 
is the diagonal degree matrix of the graph.
Finally, a node-indexed real-valued graph signal is a function $x: \mathcal{V} \to \mathbb{R}$, so that we can represent a one-dimensional graph signal as $\mathbf{x} \in \mathbb{R}^{N}$.

\subsection{Reconstruction of Time-varying Graph Signals}

The sampling and recovery of graph signals are crucial tasks in GSP \cite{11,12}.
Several studies have used the smoothness assumption to address the sampling and recovery problems for static graph signals.
The notion of global smoothness was formalized using the \textit{discrete p-Dirichlet form} \cite{24} given by:
\begin{equation}
    S_p(\mathbf{x}) = \frac{1}{p} \sum_{i \in \mathcal{V}} \left[ \sum_{j \in \mathcal{N}_i} \mathbf{A}(i,j) [\mathbf{x}(j)-\mathbf{x}(i)]^2 \right]^{\frac{p}{2}},
    \label{eqn:p-dirichlet_form}
\end{equation}
where $\mathcal{N}_i$ is the set of neighbors of node $i$.
When $p = 2$, we have $S_2(\mathbf{x})$ which is known as the graph Laplacian quadratic form $S_2(\mathbf{x}) = \sum_{(i,j) \in \mathcal{E}} \mathbf{A}(i,j) [\mathbf{x}(j)-\mathbf{x}(i)]^2 = \mathbf{x}^\mathsf{T}\mathbf{Lx}$ \cite{24}.

For time-varying graph signals, some studies assumed that the temporal differences of time-varying graph signals are smooth (\cite{23,18}).
Let $\mathbf{X} = [\mathbf{x}_1, \mathbf{x}_2, \dots, \mathbf{x}_M]$ be a time-varying graph signal, where $\mathbf{x}_s \in \mathbb{R}^{N}$ is a graph signal in $G$ at time $s$.
Qiu \textit{et al.} \cite{23} defined the smoothness of $\mathbf{X}$ as:

\begin{equation}
    S_2(\mathbf{X}) = \sum_{s=1}^M 
    \mathbf{x}_s^\mathsf{T}\mathbf{Lx}_s = tr(\mathbf{X}^{\mathsf{T}}\mathbf{LX}).
    \label{eqn:smoothness}
\end{equation}

$S_2(\mathbf{X})$ only computes the summation of the individual smoothness of each graph signal $\mathbf{x}_s~\forall~s \in \{1,2,\dots,M\}$, so we do not consider any temporal information.
To address this problem, we can define the temporal difference operator $\mathbf{D}_{h}$ as follows \cite{23}:
\begin{equation}
    \mathbf{D}_{h} = 
    \begin{bmatrix} 
        -1 & & & \\ 
        1 & -1 & & \\ 
        & 1 & \ddots & \\ 
        & & \ddots & -1 \\ 
        & & & 1 
    \end{bmatrix} \in \mathbb{R}^{M \times(M-1)}.
    \label{eq:tempdefmatrix}
\end{equation}
Therefore, we have that $\mathbf{XD}_h = [\mathbf{x}_2-\mathbf{x}_1,\mathbf{x}_3-\mathbf{x}_2,\dots,\mathbf{x}_M-\mathbf{x}_{M-1}]$.
Some studies \cite{23,18} have found that 
$S_2(\mathbf{XD}_h)$ 
shows better smoothness properties than $S_2(\mathbf{X})$ in real-world time-varying data, \textit{i.e.} $\mathbf{x}_s-\mathbf{x}_{s-1}$ 
exhibits smoothness in the graph even if $\mathbf{x}_s$ is not smooth across the graph.
Qiu \textit{et.al.} \cite{23} used $S_2(\mathbf{XD}_h)$ to present a Time-varying Graph Signal Reconstruction (TGSR) method as follows:
\begin{equation}
    \min_{\mathbf{\tilde{X}}} \frac{1}{2} \Vert \mathbf{J} \circ \mathbf{\mathbf{\tilde{X}}} - \mathbf{Y} \Vert_F^2 + \frac{\upsilon}{2} tr\left((\mathbf{\tilde{X}D}_h)^{\mathsf{T}}\mathbf{L\tilde{X}D}_h\right),
    \label{eqn:qiu_reconstruction}
\end{equation}
where $\mathbf{J} \in \{0,1\}^{N \times M}$ is a sampling matrix, $\circ$ is the Hadamard product between matrices, ${\upsilon}$ is a regularization parameter, and $\mathbf{Y} \in \mathbb{R}^{N\times M}$ is the matrix of observed values.
The optimization problem in \eqref{eqn:qiu_reconstruction} has some limitations: 1) the solution of \eqref{eqn:qiu_reconstruction} could lose performance if the real-world dataset does not satisfy the smoothness prior assumption, and 2) \eqref{eqn:qiu_reconstruction} is solved with a conjugate gradient method in \cite{23}, which has a slow convergence rate because $S_2(\mathbf{\tilde{X}D}_h)$ is ill-conditioned (\cite{18}).
Our algorithm relaxes the smoothness assumption by introducing a learnable module.
Similarly, TimeGNN is fast once the GNN parameters are learned.


\begin{figure}
    \centering
    \includegraphics[width=0.45\textwidth]{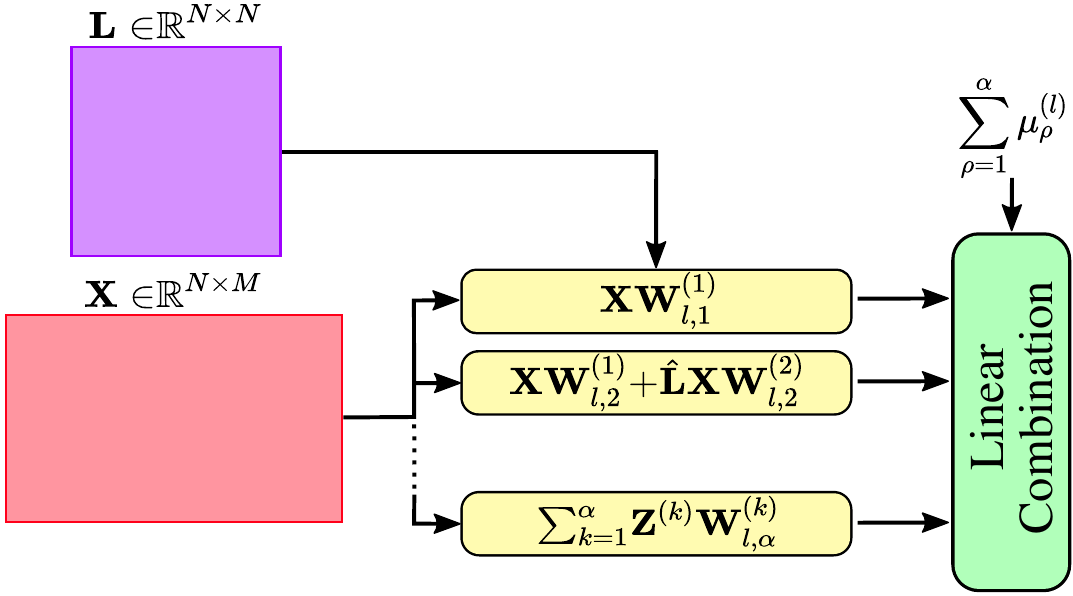}
    \caption{Cascade of Chebyshev graph convolutions.}
    \label{fig:layer}
\end{figure}

\subsection{Graph Neural Network Architecture}

\begin{figure*}
    \centering
    \includegraphics[width=0.83\textwidth]{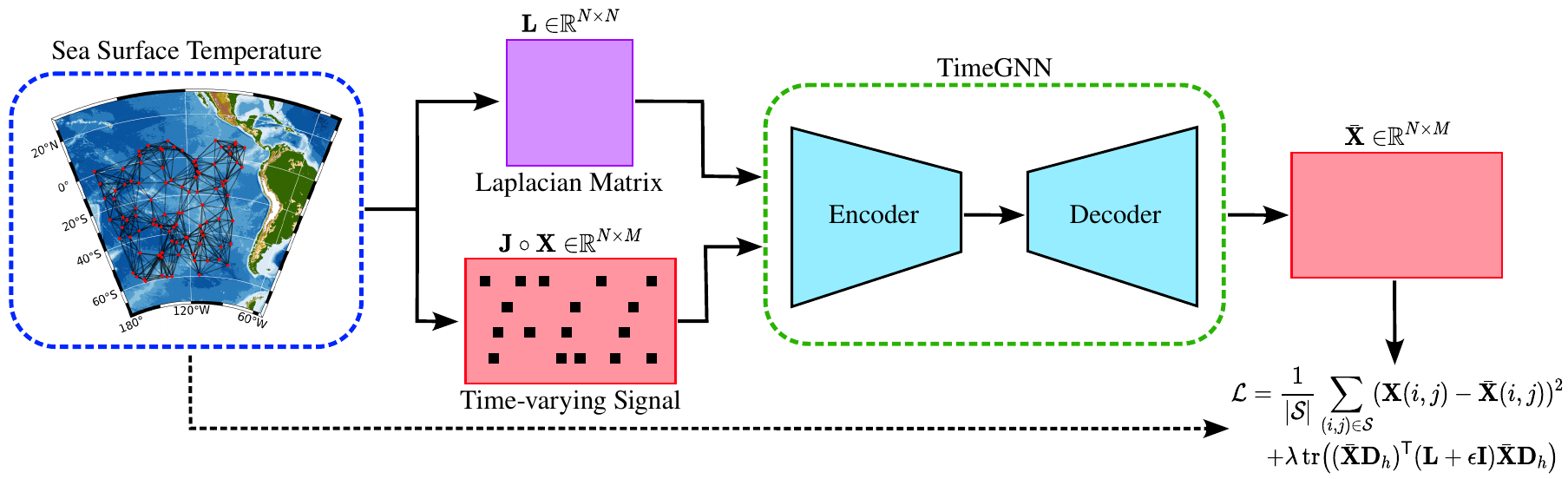}
    \caption{Pipeline of our Time Graph Neural Network (TimeGNN) for the recovery of time-varying graph signals.}
    \label{fig:pipeline}
\end{figure*}

TimeGNN is based on the Chebyshev spectral graph convolutional operator defined by Defferrard \textit{et.al.} \cite{2}, whose propagation rule is given as follows:
\begin{equation}
    \mathbf{X}' = \sum_{k=1}^K \mathbf{Z}^{(k)} \mathbf{W}^{(k)},
    \label{eqn:propagation_rule}
\end{equation}
where $\mathbf{W}^{(k)}$ is the $k$th matrix of trainable parameters, $\mathbf{Z}^{(k)}$ is computed recursively as $\mathbf{Z}^{(1)} = \mathbf{X}$, $\mathbf{Z}^{(2)} = \hat{\mathbf{L}} \mathbf{X}$, $\mathbf{Z}^{(k)} = 2 \hat{\mathbf{L}} \mathbf{Z}^{(k-1)} - \mathbf{Z}^{(k-2)}$, and $\hat{\mathbf{L}}=\frac{2\mathbf{L}}{\lambda_{\text{max}}}-\mathbf{I}$.
We use the filtering operation in \eqref{eqn:propagation_rule} to propose a new convolutional layer composed of: 1) a cascade of Chebyshev graph filters, and 2) a linear combination layer as in Fig. \ref{fig:layer}.
More precisely, we define the propagation rule of each layer of TimeGNN as follows:
\begin{equation}
    \mathbf{H}^{(l+1)} = \sum_{\rho=1}^{\alpha} \mu_{\rho}^{(l)} \sum_{k=1}^{\rho} \mathbf{Z}^{(k)} \mathbf{W}^{(k)}_{l,\rho},
    \label{eqn:propagation_rule_TimeGNN}
\end{equation}
where $\mathbf{H}^{(l+1)}$ is the output of layer $l+1$, $\alpha$ is a hyperparameter, $\mu_{\rho}^{(l)}$ is a learnable parameter, $\mathbf{Z}^{(k)}$ is recursively computed as in \eqref{eqn:propagation_rule}, and $\mathbf{W}^{(k)}_{l,\rho}$ is the $k$th learnable matrix in the layer $l$ for the branch $\rho$.
The architecture of TimeGNN is given by stacking $n$ cascade layers as in \eqref{eqn:propagation_rule_TimeGNN}, where the input is $(\mathbf{J} \circ \mathbf{X})\mathbf{D}_h$.
Finally, our loss function is such that:

\begin{align}
    \nonumber
    \mathcal{L} = \frac{1}{\vert \mathcal{S} \vert} \sum_{(i,j) \in \mathcal{S}} (\mathbf{X}(i,j) - \bar{\mathbf{X}}(i,j))^2 & \\
    \label{eqn:loss}
    + \lambda tr\left((\bar{\mathbf{X}}\mathbf{D}_h)^{\mathsf{T}}(\mathbf{L}+\epsilon \mathbf{I})\bar{\mathbf{X}}\mathbf{D}_h\right) &,
\end{align}

where $\bar{\mathbf{X}}$ is the reconstructed graph signal, $\mathcal{S}$ is the training set, with $\mathcal{S}$ a subset of the spatio-temporal sampled indexes given by $\mathbf{J}$, and $\epsilon \in \mathbb{R}^+$ is a hyperparameter.
The term $tr\left((\bar{\mathbf{X}}\mathbf{D}_h)^{\mathsf{T}}(\mathbf{L}+\epsilon \mathbf{I})\bar{\mathbf{X}}\mathbf{D}_h\right)$ is the Sobolev smoothness (\cite{18}).

We can think of TimeGNN as an encoder-decoder network with a loss function given by an MSE term plus a Sobolev smoothness regularization.
The first layers of TimeGNN encode the term $(\mathbf{J} \circ \mathbf{X})\mathbf{D}_h$ to an $H$-dimensional latent vector that is then decoded with the final layer.
As a result, we capture the spatio-temporal information using the GNN, the temporal encoding-decoding structure, and the regularization term $tr\left((\bar{\mathbf{X}}\mathbf{D}_h)^{\mathsf{T}}(\mathbf{L}+ \epsilon \mathbf{I})\bar{\mathbf{X}}\mathbf{D}_h\right)$ where we use the temporal operator $\mathbf{D}_h$.
The parameter $\lambda$ in \eqref{eqn:loss} weighs the importance of the regularization term against the MSE loss.
Figure \ref{fig:pipeline} shows the pipeline of our TimeGNN applied to a graph of the sea surface temperature in the Pacific Ocean.

%% file: tex/experiements.tex
We compare TimeGNN with Graph Convolutional Networks (GCN) \cite{3}, Natural Neighbor Interpolation (NNI) \cite{25}, TGSR \cite{23}, and Time-varying Graph signal Reconstruction via Sobolev Smoothness (GraphTRSS) \cite{18}.


\subsection{Implementation Details}

We implement TimeGNN and GCN using PyTorch and PyG \cite{26}.
We define the space search for the hyperparameters tuning of TimeGNN as follows: 1) number of layers $\{1,2,3\}$, 2) hidden units $\{2,3,\dots,10\}$, 3) learning rate $[0.005, 0.05]$, 4) weight decay $[1e-5, 1e-3]$, 5) $\lambda \in [1e-6,1e-3]$, 6) $\alpha \in \{2,3,4\}$.
Similarly, we set the following hyperparameters: 1) $\epsilon=0.05$, and 2) the number of epochs to $5,000$.
The graphs are constructed based on the coordinate locations of the nodes in each dataset with a $k$-Nearest Neighbors ($k$-NN) algorithm as in \cite{18}.
NNI, TGRS, and GraphTRSS are implemented using the code in \cite{18} in MATLAB\textsuperscript{\tiny\textregistered} 2022b.
The hyperparameters of the baseline methods are optimized following the same strategy as with TimeGNN.

\begin{figure*}
    \centering
    \includegraphics[width=0.95\textwidth]{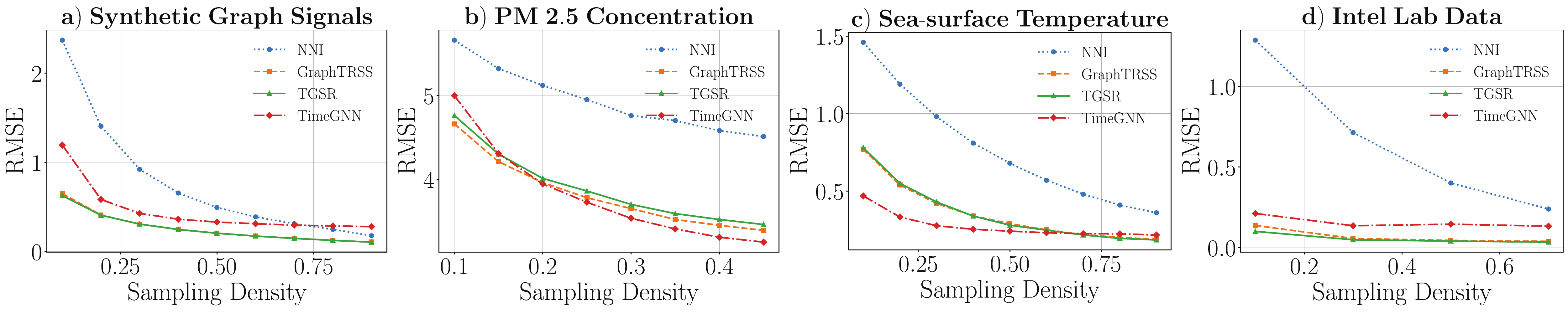}
    \caption{Comparison of TimeGNN to baseline methods in one synthetic and three real-world datasets (RMSE).}
    \label{fig:results}
\end{figure*}

\begin{table*}[]
\centering
\caption{Quantitative comparison of TimeGNN with the baselines in all datasets using the average error metrics.}
\label{tbl:summary_results}
\begin{threeparttable}
\makebox[\linewidth]{
\scalebox{0.84}{
\begin{tabular}{l|ccc|ccc|ccc|ccc}
\toprule
\multirow{2}{*}{Method} 
 & \multicolumn{3}{c|}{Synthetic Graph Signals} & \multicolumn{3}{c|}{PM2.5 Concentration} & \multicolumn{3}{c|}{Sea-surface Temperature} & \multicolumn{3}{c}{Intel Lab Data} \\
 & RMSE & MAE & MAPE & RMSE & MAE & MAPE & RMSE & MAE & MAPE & RMSE & MAE & MAPE \\
\midrule
\small GCN (Kipf and Welling \cite{3}) & $11.296$ & $8.446$ & $1.123$ & $4.657$ & $2.959$ & $0.550$ & $3.766$ & $2.922$ & $0.548$ & $2.998$ & $2.327$ & $0.120$ \\
\small NNI (Kiani \textit{et. al.} \cite{25}) & $0.775$ & $0.436$ & $0.255$ & $4.944$ & $2.956$ & $0.593$ & $0.772$ & $0.561$ & $0.067$ & $0.661$ & $0.291$ & $0.015$ \\
\small GraphTRSS (Giraldo \textit{et. al.} \cite{18}) & \color{red} $\textbf{0.260}$ & \color{blue} $\textbf{\textit{\underline{0.256}}}$ & \color{blue} $\textbf{\textit{\underline{0.178}}}$ & \color{blue} $\textbf{\textit{\underline{3.824}}}$ & \color{blue} $\textbf{\textit{\underline{2.204}}}$ & \color{blue} $\textbf{\textit{\underline{0.377}}}$ & $\color{blue}\textbf{\textit{\underline{0.357}}}$ & $\color{blue}\textbf{\textit{\underline{0.260}}}$ & $\color{blue}\textbf{\textit{\underline{0.029}}}$ & \color{red} $\textbf{0.056}$ & \color{red} $\textbf{0.023}$ & \color{red} $\textbf{0.001}$ \\
\small TGSR (Qiu \textit{et. al.} \cite{23}) & \color{blue} $\textbf{\textit{\underline{0.263}}}$ & \color{red} $\textbf{0.193}$ & \color{red} $\textbf{0.144}$ & $3.898$ & $2.279$ & $0.394$ & $0.360$ & $0.263$ & $0.030$ & \color{blue} $\textbf{\textit{\underline{0.069}}}$ & \color{blue} $\textbf{\textit{\underline{0.037}}}$ & \color{blue} $\textbf{\textit{\underline{0.002}}}$ \\
\midrule
\small TimeGNN (ours) & $0.452$ & $0.323$ & $0.226$ & \color{red} $\textbf{3.809}$ & \color{red} $\textbf{2.172}$ & \color{red} $\textbf{0.362}$ & \color{red} $\textbf{0.275}$ & \color{red} $\textbf{{0.203}}$ & \color{red} $\textbf{0.023}$ & $0.156$ & $0.095$ & $0.005$ \\
\bottomrule
\end{tabular}
}
}
\begin{tablenotes}\footnotesize
\item \scriptsize The best and second-best performing methods on each dataset are shown in {\color{red}\textbf{red}} and {\color{blue}\textbf{\textit{\underline{blue}}}}, respectively.
\end{tablenotes}
\end{threeparttable}
\end{table*}

\subsection{Datasets}


\noindent \textbf{Synthetic Graph and Signals:} We use the synthetic graph dataset developed in \cite{23}.
The graph contains $100$ nodes randomly generated from a uniform distribution in a $100 \times 100$ square area using $k$-NN.
The graph signals are generated with the recursive function $\mathbf{x}_t = \mathbf{x}_{t-1} + \mathbf{L}^{-1/2}\mathbf{f}_t$, where $\mathbf{x}_1$ is a low frequency graph signal with energy $10^4$, $\mathbf{L}^{-1/2} = \mathbf{U \mathbf{\lambda}}^{-1/2} \mathbf{U}^{\mathsf{T}}$, where $\mathbf{U}$ is the matrix of eigenvectors, $\mathbf{\lambda} = \diag(\lambda_1, \lambda_2,\ldots,\lambda_N)$ is the matrix of eigenvalues, $\mathbf{\mathbf{\lambda}}^{-1/2} = \diag(0, \lambda_2^{-1/2},\ldots,\lambda_N^{-1/2})$, and $\mathbf{f}_t$ is an i.i.d. Gaussian signal.

\noindent \textbf{PM 2.5 Concentration:} We use the daily mean concentration of PM 2.5 in the air in California, USA\footnote{\url{https://www.epa.gov/outdoor-air-quality-data}}.
Data were collected from $93$ sensors over $220$ days in 2015. 

\noindent \textbf{Sea-surface Temperature:} We use the sea-surface temperature data, which are measured monthly and released by the NOAA PSL\footnote{\url{https://psl.noaa.gov}}.
We use a sample of $100$ locations in the Pacific Ocean over a duration of $600$ months.

\noindent \textbf{Intel Lab Data:} We use the data captured by the $54$ sensors deployed at the Intel Berkeley Research Laboratory \footnote{\url{http://db.csail.mit.edu/labdata/labdata.html}}.
The data consists of temperature readings between February 28th and April 5th, 2004.

\subsection{Evaluation Metrics}

We use the Root Mean Square Error (RMSE), Mean Absolute Error (MAE), and Mean Absolute Percentage Error (MAPE) metrics, as defined in \cite{18}, to evaluate our algorithm.

\subsection{Experiments}

We construct the graphs using $k$-NN with the coordinate locations of the nodes in each dataset with a Gaussian kernel as in \cite{18}.
We follow a random sampling strategy in all experiments.
Therefore, we compute the reconstruction error metrics on the non-sampled vertices for a set of sampling densities.
We evaluate all the methods with a Monte Carlo cross-validation with $50$ repetitions for each sampling density.
For the synthetic data, $k=5$ in the $k$-NN, and the sampling densities are given by $\{0.1, 0.2, \ldots, 0.9\}$.
For PM2.5 concentration, $k=5$ and the sampling densities are $\{0.1, 0.15, 0.2,\ldots,0.45\}$.
For the sea-surface temperature, we keep $k=5$ and the sampling densities are set to $\{0.1, 0.2, \ldots, 0.9\}$.
For Intel Lab data, we set $k = 3$ and the sampling densities at $\{0.1, 0.3, 0.5, 0.7\}$.

\subsection{Results and Discussion}

Figure \ref{fig:results} shows the performance of TimeGNN against the previous methods for all datasets using RMSE.
Furthermore, Table \ref{tbl:summary_results} shows the quantitative comparisons using the averages of all metrics along the set of sampling densities.
We do not plot the performance of GCN in Fig. \ref{fig:results} because this network performs considerably worse than the other methods, as shown in Table \ref{tbl:summary_results}.
GCN was implemented using the same input and loss function as in TimeGNN.
Our algorithm outperforms previous methods for several metrics in PM2.5 concentration and sea-surface temperature datasets.
The synthetic data were created to satisfy the conditions of smoothly evolving graph signals (Definition 1 in \cite{23}), while here, we relaxed that prior assumption by adding a trainable GNN module.
Therefore, TGRS and GraphTRSS are better suited for that artificial dataset, as shown in Fig. \ref{fig:results} and Table \ref{tbl:summary_results}.
Similarly, the Intel Lab dataset is highly smooth.
Some of the reasons behind our model's success in real-world datasets are: 1) its ability to capture spatio-temporal information, 2) its encoding-decoding structure, and 3) its powerful learning module given by a cascade of Chebyshev graph convolutions.


%% file: tex/conclusions.tex
In this paper, we introduced a GNN architecture named TimeGNN for the recovery of time-varying graph signals from their samples.
Similarly, we proposed a new convolutional layer composed of a cascade of Chebyshev graph filters.
TimeGNN includes a learning module that relaxes the requirement of strict smoothness assumptions.
We found that our framework shows competitive performance against several approaches in the literature for reconstructing graph signals, delivering better performance in real datasets.
Our algorithm could help solve problems like recovering missing data from sensor networks, forecasting weather conditions, intelligent transportation systems, and many others. 

For future work, we plan to extend our framework to other graph filters like transformers \cite{27}, and alternative compact operators as introduced in \cite{28}.
Similarly, we will explore TimeGNN in highly dynamic 4D real datasets \cite{29,30}.\\

\noindent \textbf{Acknowledgments:} This work was supported by DATAIA institute as part of the ``Programme d'Investissement d'Avenir'', (ANR-17-CONV-0003) operated by CentraleSupélec, by ANR (French National Research Agency) under the JCJC project GraphIA (ANR-20-CE23-0009-01), and by the Office of Naval Research, ONR (Grant No. N00014-21-1-2760).